\documentclass[pra,twocolumn,floatfix,a4paper,superscriptaddress]{revtex4-1}
\usepackage{bm,color,graphicx,amsmath,txfonts}

\usepackage[colorlinks=true, allcolors=blue]{hyperref} 
\usepackage{soul}

\newcommand{\CC}{{\cal C}}
\newcommand{\DD}{{\cal D}}

\newcommand{\KK}{{\cal K}}

\newcommand{\MM}{{\cal M}}
\newcommand{\NN}{{\cal N}}
\newcommand{\PP}{{\cal P}}

\begin{document}

\title{Entangling the vibrational modes of two massive ferromagnetic spheres\\ using cavity magnomechanics}\thanks{This work was published in \href{https://iopscience.iop.org/article/10.1088/2058-9565/abd982}{Quantum Sci. Technol. \textbf{6}, 024005 (2021)}. The source data for the figures is available at \href{https://doi.org/10.5281/zenodo.4446839}{10.5281/zenodo.4446839}.}

\author{Jie Li}
\email{jieli6677@hotmail.com}
\affiliation{Kavli Institute of Nanoscience, Department of Quantum Nanoscience, Delft University of Technology, 2628CJ Delft, The Netherlands}
\affiliation{Zhejiang Province Key Laboratory of Quantum Technology and Device, Department of Physics and State Key Laboratory of Modern Optical Instrumentation, Zhejiang University, Hangzhou 310027, China}

\author{Simon Gr\"oblacher}
\email{s.groeblacher@tudelft.nl}
\affiliation{Kavli Institute of Nanoscience, Department of Quantum Nanoscience, Delft University of Technology, 2628CJ Delft, The Netherlands}

\begin{abstract}
We present a scheme to entangle the vibrational phonon modes of two massive ferromagnetic spheres in a dual-cavity magnomechanical system. In each cavity, a microwave cavity mode couples to a magnon mode (spin wave) via the magnetic dipole interaction, and the latter further couples to a deformation phonon mode of the ferromagnetic sphere via a nonlinear magnetostrictive interaction. We show that by directly driving the magnon mode with a red-detuned microwave field to activate the magnomechanical anti-Stokes process a cavity-magnon-phonon state-swap interaction can be realized. Therefore, if the two cavities are further driven by a two-mode squeezed vacuum field, the quantum correlation of the driving fields is successively transferred to the two magnon modes and subsequently to the two phonon modes, i.e., the two ferromagnetic spheres become remotely entangled. Our work demonstrates that cavity magnomechanical systems allow to prepare quantum entangled states at a more massive scale than currently possible with other schemes.
\end{abstract}

\maketitle

\section{Introduction}

Preparing entangled states of macroscopic, massive objects is of significance to many fundamental studies, e.g., probing the boundary between the quantum and classical worlds~\cite{Leggett,Chen,Gisin}, tests of decoherence theories at the macro scale~\cite{Bassi,Jie17,DB}, and gravitational quantum physics~\cite{VV}, among many others. Over the past decade, significant progress has been made in the field of cavity optomechanics~\cite{omRMP} in preparing entangled states of massive objects, with experimental realizations of entanglement between a mechanical oscillator and an electromagnetic field~\cite{enOM1,enOM2}, as well as between two mechanical oscillators~\cite{enMM1,enMM2,enMM3}. All those entangled states were created and detected by utilizing the radiation pressure interaction, or, more specifically, the optomechanical two-mode squeezing and beamsplitter (state-swap) interactions, realized by driving the cavity with a blue- and red-detuned electromagnetic field, respectively, and optimally working in the resolved sideband limit.

In analogy to cavity optomechanics, in recent years cavity magnomechanics (CMM)~\cite{Tang16} has received increasing attention, owing to its potential for realizing quantum states at a more macroscopic scale~\cite{Jie18,Jie19,Jie19b} and possible applications in quantum information processing and quantum sensing~\cite{NakaRev}. In these systems, a magnon mode (spin wave) of a ferromagnetic yttrium-iron-garnet (YIG) sphere couples to a microwave (MW) cavity field~\cite{S1,S2,S3,S4,S5,S6}, and simultaneously couples to the vibrational phonon mode (deformation mode) of the sphere via the magnetostrictive force~\cite{Kittel58}. Owing to the high spin density and the low damping rate of YIG, the interaction between the MW cavity field and the magnon mode can easily enter the strong coupling regime~\cite{S1,S2,S3,S4,S5,S6}, thus providing an excellent platform for the study of strong interaction between light and matter. Many interesting phenomena have been explored in the context of cavity magnonics, such as a magnon gradient memory~\cite{TangNC}, exceptional points~\cite{YouNC}, the manipulation of distant spin currents~\cite{spinCur}, level attraction~\cite{Hu18}, nonreciprocity~\cite{Hu19}, among others. In the tripartite system of CMM, the phonon mode is typically of low frequency due to the large size of the sphere. The magnomechanical interaction is a radiation pressure-like, dispersive interaction~\cite{Tang16,Oriol} and the Hamiltonian is given by $H/\hbar =G_0 m^{\dag} m (b+b^{\dag})$, where $m$ ($b$) is the annihilation operator for the magnon (phonon) mode, and $G_0$ is the single-magnon magnomechanical coupling rate. The fact that this Hamiltonian takes the same form as that of the optomechanical interaction allows us to predict new phenomena in CMM from known results in cavity optomechanics.

To date, magnomechanically induced transparency (MMIT) has been experimentally observed~\cite{Tang16}, and multi-window MMIT has been proposed by coupling a cavity mode to two YIG spheres~\cite{MMIT}. Quantum effects in CMM have been first studied in Ref.~\cite{Jie18}, which shows the possibility of creating genuine tripartite magnon-photon-phonon entanglement and cooling of the mechanical motion. Furthermore, proposals have been made for generating squeezed vacuum states of magnons and phonons~\cite{Jie19}, and entangled states of two magnon modes in CMM~\cite{Jie19b}. Quite recently, CMM has been used to produce stationary entangled MW fields by coupling a magnon mode to two MW cavities~\cite{Jie20}. These protocols~\cite{Jie18,Jie19,Jie19b,Jie20} essentially utilize the nonlinear magnetostrictive interaction effectively activated by properly driving the magnon mode with a magnetic field, which can be experimentally realized by directly driving the YIG sphere with a small MW loop antenna~\cite{Wang}, allowing to implement the magnomechanical beamsplitter or two-mode squeezing interactions. Other quantum effects like tripartite Einstein-Podolsky-Rosen steering have also been studied~\cite{Tan}. In addition, many other interesting topics have been explored in CMM, including magnetically tunable slow light~\cite{Xiong}, phonon lasing~\cite{LiC}, thermometry~\cite{Davis}, and parity-time-related phenomena~\cite{Liu,Sun,HongFu}.

\begin{figure}[t]
\includegraphics[width=\linewidth]{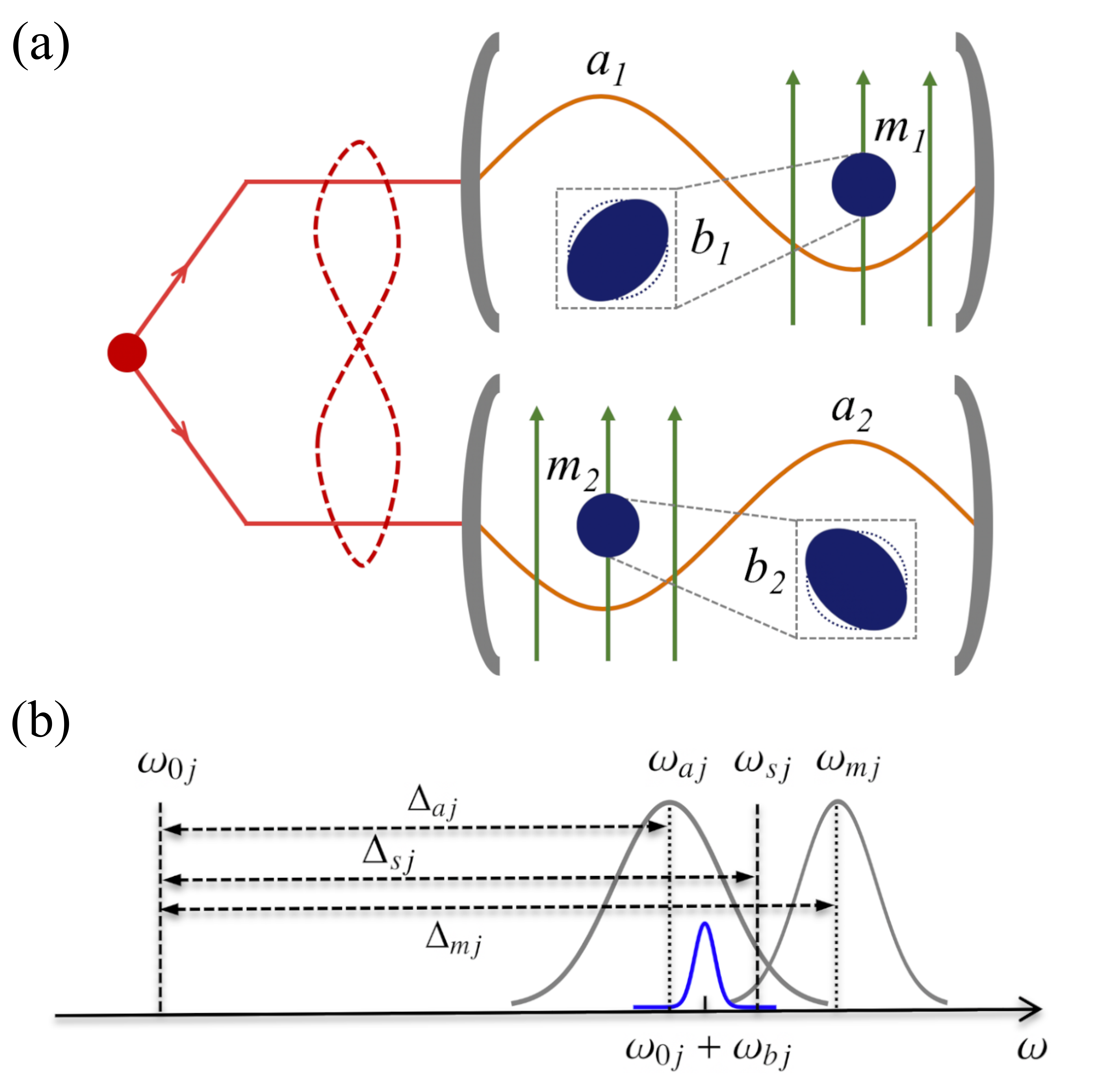}
\caption{(a) Two YIG spheres are placed inside two MW cavities, which are driven by a two-mode squeezed vacuum MW field. Each sphere is placed in a uniform bias magnetic field and near the maximum magnetic field of the cavity mode, and is directly driven by a strong red-detuned MW field (not shown) to enhance magnon-phonon coupling. (b) The frequencies of the modes and drive fields in the cavities ($j{=}1,2$) are shown. The MW cavity with resonance frequency $\omega_{aj}$ is driven by the $j$th mode (with frequency $\omega_{sj}$) of the two-mode squeezed MW field. The magnon mode with frequency $\omega_{mj}$ is driven by another strong red-detuned MW field of frequency $\omega_{0j}$. The mechanical motion of frequency $\omega_{bj}$ scatters the driving photons onto two sidebands at frequencies $\omega_{0j}\pm \omega_{bj}$. For the case when the cavity mode, magnon mode, and the squeezed drive field are resonant with the blue mechanical sideband within each cavity, the two phonon modes of the two independent, spatially separated YIG spheres become entangled.}
\label{Fig1}
\end{figure}

In this article, we present the first proposal to entangle the vibrational phonon modes of two massive YIG spheres. We would like to note that the entanglement of two magnon modes~\cite{Jie19b,Zhedong,Yung,prb,Jaya,Yu,Qian} is a non-classical state of a large number of spins inside the YIG spheres. In contrast, here we consider the entanglement of the vibrational modes of the whole spheres. The phonon mode typically has a much lower frequency than the magnon mode~\cite{Tang16,Jie18,Jie19,Jie19b,NakaRev,S1,S2,S3,S4,S5,S6} (MHz vs.\ GHz), indicating increasing susceptibility to the thermal noise from the surrounding environment, which significantly increases the difficulty to prepare phonon entangled states. The system consists of two MW cavities each containing a YIG sphere which supports a magnon mode and a deformation phonon mode. The two cavities are driven by a two-mode squeezed vacuum MW field, which entangles the two MW intra-cavity fields, and, owing to the cavity-magnon beamsplitter interaction, the two magnon modes thus get entangled. We then directly drive each magnon mode with a strong red-detuned MW field, activating the magnomechanical state-swap interaction allowing for the transfer of squeezing from the magnon mode to the phonon mode. Therefore, the two phonon modes of two YIG spheres become entangled. Similar ideas of transferring an entangled state from light to macroscopic mechanical oscillators have been provided for optomechanical systems~\cite{Zhang03,Mauro1,Mauro2}.

\section{The model}

We consider a dual-cavity magnomechanical system, with each cavity containing a MW, a magnon and a phonon mode, as depicted in Fig.~\ref{Fig1}. The magnon and phonon modes are supported by the YIG sphere, which has a typical diameter in the 100~$\mu$m range~\cite{Tang16}. The magnon mode is embodied by the collective motion of a large number of spins in the YIG sphere, and the phonon mode is the deformation mode of the sphere caused by the magnetostrictive force~\cite{Kittel58}. In each cavity, the magnon mode couples to the MW cavity mode via the magnetic dipole interaction, and to the phonon mode via the nonlinear radiation pressure-like magnomechanical interaction. In our scheme, each magnon mode is directly driven through a strong red-detuned MW field, realized by, e.g., driving the YIG sphere with a small loop antenna at the end of a superconducting MW line~\cite{Wang,Jie20}, which enhances the magnomechanical coupling strength, cools the phonon mode~\cite{Jie18}, and activates the magnon-phonon state-swap interaction~\cite{Jie19}. The Hamiltonian of the system is given by
\begin{equation}\label{Hamilt}
\begin{split}
{\cal H}/\hbar = \!\! \sum_{j=1,2} & \bigg \{  \omega_{aj} a_j^{\dag} a_j \,{+}\, \omega_{mj} m_j^{\dag} m_j \,{+}\, \omega_{bj} b_j^{\dag} b_j \,{+}\,  g_j (a_j^{\dag} m_j \,{+}\, a_j m_j^{\dag})  \\
&+ G_{0j} m_j^{\dag} m_j (b_j^{\dag} \,{+}\, b_j) + i \Omega_j (m_j^{\dag} e^{-i \omega_{0j} t} \,{-}\, m_j e^{i \omega_{0j} t})  \bigg \}  ,
\end{split}
\end{equation}
where $a_j$, $m_j$, and $b_j$ ($\omega_{aj}$, $\omega_{mj}$, and $\omega_{bj}$) are the annihilation operators (resonance frequencies) of the cavity, magnon and phonon modes, respectively, satisfying $[O_j, O_j^{\dag}]=1$ $(O=a, m,b)$, with $j=1,2$. The magnon frequency $\omega_{mj}$ can be adjusted by varying the external bias magnetic field $H_j$ via $\omega_{mj} \,{=}\, \gamma_0 H_j$, where the gyromagnetic ratio for YIG $\gamma_0/2\pi \,{=}\, 28$ GHz/T. $g_j$ is the cavity-magnon coupling rate, which can be much larger than the dissipation rates of the two modes, $g_j > \kappa_{a_j}, \kappa_{m_j}$~\cite{S1,S2,S3,S4,S5,S6}. $G_{0j}$ is the bare magnon-phonon coupling rate, which is usually quite small, but can be enhanced by driving the magnon mode with a strong MW field. The Rabi frequency $\Omega_j=\frac{\sqrt{5}}{4} \gamma_0 \! \sqrt{N_j} B_{0j}$~\cite{Jie18} denotes the coupling rate between the magnon mode and its driving magnetic field with frequency $\omega_{0j}$ and amplitude $B_{0j}$, while $N_j \,{=}\, \rho V_j$ is the total number of spins, with $\rho=4.22 \times 10^{27}$ m$^{-3}$ the spin density of YIG and $V_j$ is the volume of the spheres. Note that for the magnon modes, we have expressed the collective spin operators in terms of Boson (oscillator) operators via the Holstein-Primakoff transformation~\cite{HPT} under the condition of low-lying excitations, $\langle m_j^{\dag} m_j \rangle \,{\ll} \, 2N s$ (for simplicity we assume the two spheres to be of the same size and thus of the same total number of spins $N$), where $s\,{=}\,\frac{5}{2}$ is the spin number of the ground state Fe$^{3+}$ ion in YIG.

We now assume the two cavities to be driven by a continuous, two-mode squeezed vacuum MW input field with frequency $\omega_{sj}$ and each cavity to be resonant with the squeezed drive as well as the magnon mode, such that $\omega_{aj}=\omega_{mj}=\omega_{sj}$, or $\Delta_{aj}=\Delta_{mj}=\Delta_{sj} \equiv \Delta_j$ ($j=1,2$), where the detunings $\Delta_{Oj}=\omega_{Oj}-\omega_{0j}$ $(O=a, m, s)$ are with respect to the magnon drive frequency $\omega_{0j}$, see Fig.~\ref{Fig1}b. This situation is easily realized as all three frequencies are tunable, and the resonant case also corresponds to the optimal situation for transferring squeezing from the driving field to the magnon mode~\cite{Jie19,Yu}. Note that $\Delta_1=\Delta_2$ is however not required as each should match the frequency of the phonon mode of the respective YIG sphere, i.e., $\Delta_j \simeq \omega_{bj}$. This corresponds to the magnon mode being resonant with the blue mechanical sideband (see Fig.~\ref{Fig1}b), which is required for realizing the magnomechanical state-swap interaction in each sphere, such that the squeezing can further be transferred from the magnon mode to the phonon mode.

The quantum Langevin equations (QLEs) for describing the cavity, magnon, and phonon modes are given by (in the frame rotating at the magnon drive frequency $\omega_{0j}$)
\begin{equation}\label{QLE1}
\begin{split}
\dot{a}_j&= - (i \Delta_j + \kappa_{a_j}) a_j - i g_j m_j + \sqrt{2 \kappa_{a_j}} a_j^{\rm in},  \\
\dot{m}_j&= - (i \Delta_j + \kappa_{m_j}) m_j - i g_j a_j - i G_{0j} m_j (b_j^{\dag} \,{+}\, b_j) + \Omega_j +\!\! \sqrt{2 \kappa_{m_j}} m_j^{\rm in},  \\
\dot{b}_j&= - (i \omega_{bj} + \gamma_j ) b_j - i G_{0j} m_j^{\dag} m_j +  \sqrt{2 \gamma_j} b_j^{\rm in} ,   \\
\end{split}
\end{equation}
where $\gamma_j$ are the mechanical damping rates, and $a_j^{\rm in}$, $m_j^{\rm in}$ and $b_j^{\rm in}$ are input noise operators for the cavity, magnon, and phonon modes, respectively. Owing to the injection of a two-mode squeezed vacuum field, which shapes the noise properties of two MW cavity fields, the input noise of the two cavities $a_{1,2}^{\rm in}$ become quantum correlated and possess the correlation functions
\begin{equation}\label{noise1}
\begin{split}
\langle a_j^{\rm in}(t) \, a_j^{\rm in \dag}(t')\rangle &= ({\cal N}{+}1) \,\delta(t{-}t'),  \\
\langle a_j^{\rm in \dag}(t) \, a_j^{\rm in}(t')\rangle &= {\cal N} \, \delta(t{-}t'),  \\
\langle a_j^{\rm in}(t) \, a_k^{\rm in}(t')\rangle &= {\cal M} \, e^{-i (\Delta_j t+ \Delta_k t')} \delta(t{-}t'),  \\ 
\langle a_j^{\rm in \dag}(t) \, a_k^{\rm in \dag}(t')\rangle &= {\cal M}^* e^{i (\Delta_j t+ \Delta_k t')}\, \delta(t{-}t'), \,\,\, (j \,{\ne}\, k \,{=}\,1,2) 
\end{split}
\end{equation}
where ${\cal N}\,\,{=}\,\, \sinh^2 r$, ${\cal M} \,\,{=}\,\, \sinh r \cosh r $. Here $r$ is the squeezing parameter of the two-mode squeezed vacuum field, which is typically produced by a Josephson parametric amplifier (JPA)~\cite{sqzMW1}, a Josephson mixer~\cite{sqzMW2}, or the combination of a JPA and a MW beamsplitter~\cite{sqzMW3,sqzMW4}. Note that the phase factors in the noise correlations are due to the non-zero frequencies of the squeezed driving fields in the reference frame. The input noise of the magnon and phonon modes $O_j^{\rm in}$ ($O=m,b$) are of zero mean value and correlated as
\begin{equation}\label{noise2}
\begin{split}
\langle  O_{j}^{\rm in}(t) O_{j}^{\rm in\dag }(t^{\prime }) \rangle &= (N_{O_{j}}+1)\delta(t-t^{\prime }),  \\
\langle  O_{j}^{\rm in\dag}(t) O_{j}^{\rm in}(t^{\prime }) \rangle &= N_{O_{j}} \delta(t-t^{\prime }),  \\
\end{split}
\end{equation}
where $N_{O_{j}}=[\exp (\frac{\hbar \omega _{O_{j}}}{k_{B}T})-1]^{-1}$ is the equilibrium mean thermal magnon/phonon number, and $k_B$ the Boltzmann constant and $T$ the bath temperature. For simplicity, we assume the two cavities to be at the same environment and thus bath temperature.

Since the magnon mode in each cavity is strongly driven, it has a large amplitude $|\langle m_j \rangle| \gg 1$, and owing to the cavity-magnon linear coupling, the cavity field also has a large amplitude $|\langle a_j \rangle| \gg 1$. This allows us to linearize the system dynamics (essentially the nonlinear magnetostrictive interaction) around the semiclassical averages by writing any operator as $O_j=\langle O_j \rangle +\delta O_j$ ($O\, {=}\, a,m,b$) and neglecting small second-order fluctuation terms. As a result, the QLEs~\eqref{QLE1} are separated into two sets of equations for semiclassical averages and for quantum fluctuations, respectively. By solving the former set of equations, we obtain the {\it steady-state} solution for the average
\begin{equation}\label{aveM}
\langle m_j \rangle =  \frac{ ( i \Delta_j +\kappa_{a_j}) \, \Omega_j  }{ g_j^2 \! + ( i \tilde{ \Delta}_j + \kappa_{m_j}) ( i \Delta_j + \kappa_{a_j}) },
\end{equation}
where $\tilde{ \Delta}_j = \Delta_j + 2 G_{0j} {\rm Re} \langle b_j \rangle$ is the effective magnon-drive detuning including the frequency shift caused by the magnomechanical interaction. This frequency shift is typically small because of a small $G_{0j}$~\cite{Tang16},  $|\tilde{ \Delta}_j - \Delta_j | \ll \Delta_j \simeq \omega_{bj}$, and thus hereafter we can safely assume $\tilde{ \Delta}_j \simeq \Delta_j$. When $\Delta_j \simeq \omega_{bj} \gg  \kappa_{a_j}, \kappa_{m_j}$, which is easily satisfied~\cite{Tang16}, Eq.~\eqref{aveM} takes a simple approximate form $\langle m_j \rangle \simeq  i  \Delta_j \Omega_j / (g_j^2 -\Delta_j^2)$, which is a pure imaginary number. The solutions of $\langle a_j \rangle$ and $\langle b_j \rangle$ can then be obtained by $\langle a_j \rangle = -i g_j \langle m_j \rangle/(i \Delta_j + \kappa_{a_j})$, and $\langle b_j \rangle = -i G_{0j} |\langle m_j \rangle|^2 /(i \omega_{bj} + \gamma_j) \simeq -G_{0j} |\langle m_j \rangle|^2 /\omega_{bj}$, taking into account the mechanical $Q$ factor is typically high, $\omega_{bj}/ \gamma_j \gg 1$. The average $\langle b_j \rangle$ is therefore a real number, implying that the average of mechanical momentum, $\langle p_j \rangle \,{=} \sqrt{2} \, {\rm Im} \langle b_j \rangle$, is zero in the steady state.
 
The QLEs for the quantum fluctuations are given by
\begin{equation}\label{QLE2}
\begin{split}
\delta\dot{a}_j&= - (i \Delta_j + \kappa_{a_j}) \delta a_j - i g_j \delta m_j +\! \sqrt{2 \kappa_{a_j}} a_j^{\rm in},  \\
\delta\dot{m}_j&= - (i \Delta_j + \kappa_{m_j}) \delta m_j - i g_j \delta a_j  {-} \, G_j (\delta b_j^{\dag} {+} \delta b_j ) +\!\! \sqrt{2 \kappa_{m_j}} m_j^{\rm in},  \\
\delta\dot{b}_j&= - (i \omega_{bj} + \gamma_j) \delta b_j - G_j (\delta m_j^{\dag} - \delta m_j ) +\!\! \sqrt{2 \gamma_j} b_j^{\rm in},  \\
\end{split}
\end{equation}
where $G_j \,{=}\, i G_{0j} \langle m_j \rangle$ is the effective magnomechanical coupling rate. We now move to a reference frame rotating at frequency $\Delta_j=\omega_{bj}$, by introducing the slowly moving operators $\tilde{O}$, $\delta a_j = \delta \tilde{a}_j e^{-i \Delta_j t}$, $\delta m_j = \delta \tilde{m}_j e^{-i \Delta_j t}$, and $\delta b_j = \delta \tilde{b}_j e^{-i \omega_{bj} t}$, where $\delta \tilde{a}_j$, $\delta \tilde{m}_j$, and $\delta \tilde{b}_j$ are defined in the new reference frame. We make the same transformation for the input noise operators, and obtain noise correlations in the new frame, which remain the same as in Eqs.~\eqref{noise1}-\eqref{noise2} but without the phase factors in Eq.~\eqref{noise1}, as we are now in a frame that is resonant with the squeezed drive field. By substituting the above transformations into the QLEs~\eqref{QLE2}, and neglecting fast oscillating non-resonant terms, we obtain the following QLEs 
\begin{equation}\label{QLE3}
\begin{split}
\delta\dot{\tilde a}_j&= - \kappa_{a_j} \delta \tilde{a}_j - i g_j \delta \tilde{m}_j +\! \sqrt{2 \kappa_{a_j}} \tilde{a}_j^{\rm in},  \\
\delta\dot{\tilde m}_j&= - \kappa_{m_j} \delta \tilde{m}_j - i g_j \delta \tilde{a}_j  - G_j \delta \tilde{b}_j +\! \sqrt{2 \kappa_{m_j}} \tilde{m}_j^{\rm in},  \\
\delta\dot{\tilde b}_j&= - \gamma_j \delta \tilde{b}_j + G_j  \delta \tilde{m}_j  +\! \sqrt{2 \gamma_j} \tilde{b}_j^{\rm in},  \\
\end{split}
\end{equation}
which are a good approximation if the condition $\Delta_j \,{=}\, \omega_{bj} \gg G_j, g_j, \kappa_{a_j}, \kappa_{m_j}, \gamma_j$ is satisfied. The QLEs~\eqref{QLE3} clearly reveal a beamsplitter interaction in the cavity-magnon and magnon-phonon subsystems, which allows for cooling the phonon modes and the transfer of two-mode squeezing from the driving fields to the two cavity modes, then to the two magnon modes, and finally to the two phonon modes of the two spatially separated YIG spheres.

\section{Entanglement of two YIG spheres}

We now proceed to study the entanglement of the two phonon modes. We rewrite the QLEs~\eqref{QLE3} in terms of quadrature fluctuations, which can be cast in the following form
\begin{equation}\label{uAn}
\dot{u} (t) = A u(t) + n(t) ,
\end{equation}
where $u=\big( \delta x_1, \delta y_1 , \delta x_2, \delta y_2, \delta X_1, \delta Y_1, \delta X_2, \delta Y_2, \delta q_1, \delta p_1, \delta q_2, \\\delta p_2 \big)^T$, and the quadrature fluctuation operators are defined as $\delta x_{j} \,\,{=}\,\, (\delta \tilde{a}_{j} \,\,{+}\,\, \delta \tilde{a}_{j}^{\dag })/\sqrt{2}$, $\delta y_{j} \,\,{=}\,\, i(\delta \tilde{a}_{j}^{\dag } \,\,{-}\,\, \delta \tilde{a}_{j})/\sqrt{2}$, $\delta X_j \,{=}\, (\delta \tilde{m}_{j} \,{+}\, \delta \tilde{m}_{j}^{\dag })/\sqrt{2}$, $\delta Y_j \,{=}\, i(\delta \tilde{m}_{j}^{\dag } \,{-}\, \delta \tilde{m}_{j})/\sqrt{2}$, $\delta q_j \!\!\!\! \!\! = (\delta \tilde{b}_{j} \,{+}\, \delta \tilde{b}_{j}^{\dag })/\sqrt{2}$, and $\delta p_j \,{=}\, i(\delta \tilde{b}_{j}^{\dag } \,{-}\, \delta \tilde{b}_{j})/\sqrt{2}$. Similarly, we can define the quadratures of the input noise $O_{j}^{\rm in}$ ($O\,{=}\, x,\, y,\, X,\, Y,\, q,\, p$). For simplicity, we have removed the tilde signs for the quadrature operators. $n = \big( \! \! \sqrt{2 \kappa_{a_1}} x_1^{\rm in}, \sqrt{2 \kappa_{a_1}} y_1^{\rm in}, \sqrt{2 \kappa_{a_2}} x_2^{\rm in}, \sqrt{2 \kappa_{a_2}} y_2^{\rm in},  \sqrt{2 \kappa_{m_1}} X_1^{\rm in}, \sqrt{2 \kappa_{m_1}} Y_1^{\rm in}, \\  \sqrt{2 \kappa_{m_2}} X_2^{\rm in}, \sqrt{2 \kappa_{m_2}} Y_2^{\rm in},  \sqrt{2 \gamma_1} q_1^{\rm in}, \sqrt{2 \gamma_1} p_1^{\rm in}, \sqrt{2 \gamma_2} q_2^{\rm in}, \sqrt{2 \gamma_2} p_2^{\rm in}  \big)^T$ is the vector of input noise, and the drift matrix $A$ is large and its specific form is provided in Appendix A. 
    
Owing to the linearized dynamics and the Gaussian nature of input noise, the system preserves Gaussian states for all times. The steady state of the system is a six-mode Gaussian state, which is fully characterized by a $12 \times 12$ covariance matrix (CM) ${\CC}$, whose entries are defined as ${\CC}_{sk}(t) = \frac{1}{2} \langle u_s(t) u_k(t') + u_k(t') u_s(t) \rangle$ $(s,k \,{=}\, 1,2,...,12)$. The stationary CM ${\CC}$ can be obtained by directly solving the Lyapunov equation~\cite{DV07,Hahn}
\begin{equation}\label{Lyap}
A \CC+\CC A^T = -\DD,
\end{equation}
where $\DD$ is the diffusion matrix defined by $\DD_{sk} \delta (t-t') = \frac{1}{2} \langle n_s(t) n_k(t') +n_k(t') n_s(t) \rangle$. It can be written in the form of a direct sum, $\DD=\DD_a \oplus \DD_m \oplus \DD_b$, where $\DD_a$ is related to the squeezed input noise of the two cavity modes
\begin{widetext}
\begin{equation}
\DD_a=\left( 
\begin{array}{cccc}
\kappa _{a_{1}}(2\NN+1) & 0 & \sqrt{\kappa _{a_{1}}\kappa _{a_{2}}}(\MM + \MM^{\ast}) & i\sqrt{\kappa _{a_{1}}\kappa _{a_{2}}}(-\MM + \MM^{\ast }) \\ 
0 & \kappa _{a_{1}}(2\NN+1) & i\sqrt{\kappa _{a_{1}}\kappa _{a_{2}}}(-\MM +\MM^{\ast}) & -\sqrt{\kappa _{a_{1}}\kappa _{a_{2}}}(\MM + \MM^{\ast }) \\ 
\sqrt{\kappa _{a_{1}}\kappa _{a_{2}}}(\MM + \MM^{\ast }) & i\sqrt{\kappa_{a_{1}}\kappa _{a_{2}}}(- \MM + \MM^{\ast }) & \kappa _{a_{2}}(2\NN+1) & 0 \\ 
i\sqrt{\kappa _{a_{1}}\kappa _{a_{2}}}(- \MM + \MM^{\ast }) & -\sqrt{\kappa_{a_{1}}\kappa _{a_{2}}}(\MM + \MM^{\ast }) & 0 & \kappa _{a_{2}}(2\NN+1)
\end{array}
\right) , 
\end{equation}
\end{widetext}
and $\DD_{m}$ ($\DD_{b}$) is associated with the thermal input noise for two magnon (phonon) modes, $\DD_m={\rm diag} \big[ \kappa _{m_{1}}(2N_{m_{1}}+1),\kappa_{m_{1}}(2N_{m_{1}}+1),\kappa_{m_{2}}(2N_{m_{2}}+1),\kappa_{m_{2}}(2N_{m_{2}}+1) \big]$, and $\DD_b={\rm diag} \big[ \gamma_1 (2N_{b_1}+1), \gamma_1 (2N_{b_1}+1), \gamma_2 (2N_{b_2}+1), \gamma_2 (2N_{b_2}+1) \big]$. Once the CM of the system is obtained, one can then extract the state of the two phonon modes and calculate their entanglement property. We adopt the logarithmic negativity~\cite{LogNeg} to quantify the entanglement of the Gaussian states, whose definition is provided in Appendix B.

\begin{figure}[b]
\hskip-0.15cm\includegraphics[width=\linewidth]{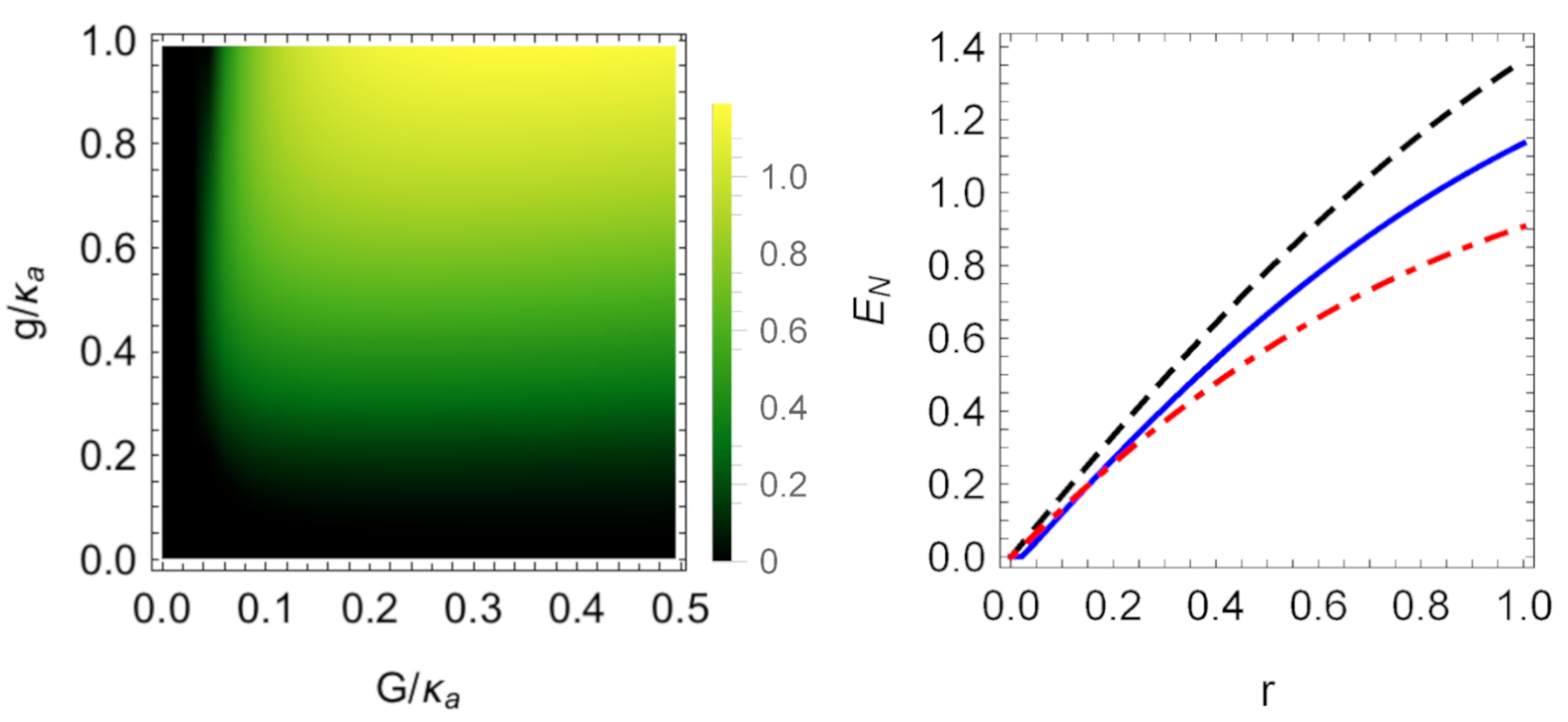}
\caption{(a) Entanglement (logarithmic negativity) $E_N$ of the two phonon modes as a function of the two coupling rates $G$ and $g$ for a two-mode squeezed driving field with $r=1$. (b) Stationary cavity-cavity (black dashed), magnon-magnon (red dotted-dashed), and phonon-phonon (blue solid) entanglement vs.\ $r$, with $G = 0.2\kappa_a$ and $g = \kappa_a$. All other parameters are taken from~\cite{Tang16} and are given in the text.}
\label{Fig2}
\end{figure}

We present our main result of the steady-state entanglement between two YIG spheres in Fig.~\ref{Fig2}. The stability is guaranteed by the negative eigenvalues (real parts) of the drift matrix $A$. We have adopted experimentally feasible parameters~\cite{Tang16}:\ $\omega_{a} \,\,{=}\,\, \omega_{m} \,\,{=}\,\, \omega_{s} \,\,{=}\, 2\pi \times 10$~GHz, $\omega_{b_1} = 2\pi \times 10$~MHz, $\omega_{b_2} = 1.2\omega_{b_1}$, $\gamma = 2\pi \times 100$~Hz, $\kappa_{a} \,\,{=}\,\, 2\pi \times 3$~MHz, $\kappa_{m} \,\,{=}\,\, \kappa_{a}/5$, and $T \,\,{=}\,\, 10$~mK. Note that, in our model the linewidth of the magnon (cavity) mode is defined as $2\kappa_m$ ($2\kappa_a$). Here we take $2\kappa_m\,\,{=}\,\,1.2$~MHz, which is larger than the magnon intrinsic dissipation (typically of the order of 1~MHz), as well as the demonstrated value $1.12$~MHz~\cite{Tang16}. For simplicity, we have assumed equal frequencies for the two cavity (magnon) modes, and squeezed driving fields, $\omega_{O_1} \,{=}\,\, \omega_{O_2} \, {\equiv} \,\, \omega_O$ ($O\,{=}\,a,m,s$)~\cite{Note}, due to their flexible tunability, but generally different frequencies for the two phonon modes. This means that the frequencies of the two magnon drive fields are also different because $\omega_{0j} = \omega_m - \omega_{bj}$. For convenience, we have also assumed equal dissipation rates for all pairs of modes of the same type. In Fig.~\ref{Fig2}a, we show the mechanical entanglement versus two coupling rates $g_1 = g_2 \equiv g$ and $G_1 = G_2 \equiv G$, and consider $g,G \le \kappa_a \ll \omega_{b_{1,2}}$, in order to meet the condition used for deriving Eq.~\eqref{QLE3}. Figure~\ref{Fig2}b shows that in the steady state the two cavity/magnon/phonon modes are all entangled, and the entanglement increases with larger $r$. The mechanical entanglement is even stronger than the magnon entanglement when $r\,\,{>}\,{\sim}\,0.2$, although the former is transferred from the latter. This is possible because the cavities are continuously driven, and the total entanglement is distributed among the three different subsystems with steady-state bipartite entanglement. We use a relatively larger cavity decay rate $\kappa_{a} \gg \kappa_{m}$, which has been shown to be an optimal condition for obtaining magnon entanglement~\cite{Yu}, which is a pre-requisite for phonon entanglement in our protocol. We would like to note that, for the parameters of Fig.~\ref{Fig2}b, the entanglement of any two modes of different types are either negligibly small or zero.

\begin{figure}
\hskip-0.4cm\includegraphics[width=0.8\linewidth]{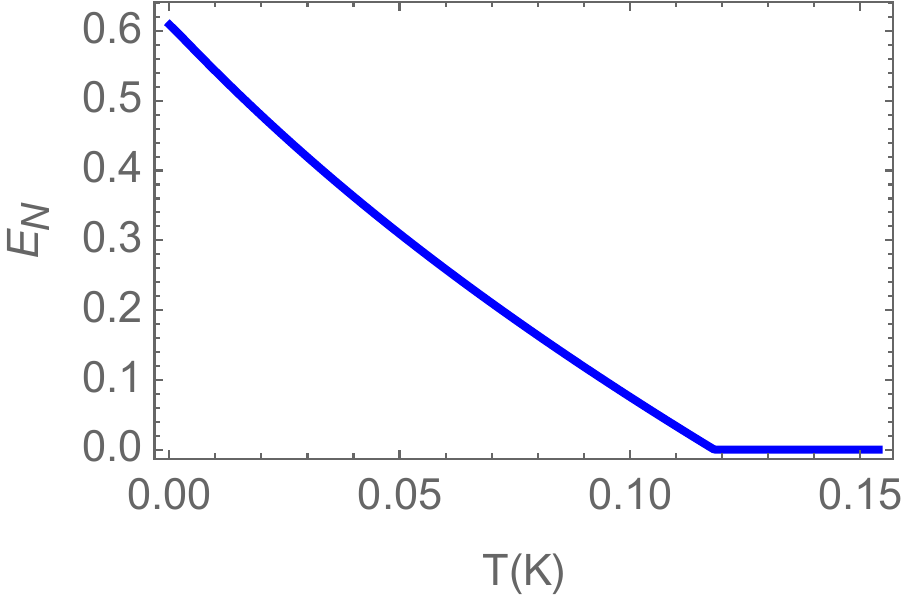}
\caption{Calculation of the steady-state mechanical entanglement $E_N$ vs.\ bath temperature $T$ with $r=0.4$, clearly showing that the non-classical correlation between the two phonon modes survives up to the temperature of $118$~mK. All other parameters are the same as in Fig.~\ref{Fig2}b.}
\label{Fig3}
\end{figure}

In Fig.~\ref{Fig3}, we show the entanglement as a function of bath temperature for a two-mode squeezed vacuum of $r=0.4$. This corresponds to a logarithmic negativity $E_N=0.8$~\cite{Note2} of the driving field, which has been experimentally demonstrated in Ref.~\cite{sqzMW2}. With such a driving field, we obtain mechanical entanglement $E_N=0.54$ for $T=10$~mK, and the entanglement survives up to 118 mK.

Lastly, we would like to discuss how to detect the entanglement. The generated entanglement of two YIG spheres can be verified by measuring the CM of the two phonon modes~\cite{enMM2,enMM3}. The mechanical quadratures can be measured by coupling each sphere to an additional optical cavity which is driven by a {\it weak} red-detuned laser. This yields an optomechanical state-swap interaction which maps the phonon state onto the cavity output field~\cite{JieSimon}. By homodyning this field, the mechanical quadratures can be measured, based on which the CM can be reconstructed.

\section{Validity of the model} 

We now discuss the validity of the approximations that were made in our model. For the magnon modes, we have assumed low-lying excitations, $\langle m_j^{\dag} m_j \rangle \,{\ll} \, 2N s$, in order to express the collective spin operators in terms of Boson operators. For a 250-$\mu$m-diameter YIG sphere, $N \,\,{\simeq}\,\, 3.5 \times 10^{16}$, and the coupling $G=0.2 \kappa_a = 2\pi \times 0.6$~MHz used in Fig.~\ref{Fig2}b and Fig.~\ref{Fig3} corresponds to $|\langle m \rangle| \simeq 1.2 \times 10^7$ for $G_0/2\pi = 50$~mHz. Therefore, $\langle m^{\dag} m \rangle \simeq 1.4 \times 10^{14} \ll 2N s=1.7 \times 10^{17}$, which is well satisfied. 

We have also assumed the magnon frequency shift caused by the magnomechanical interaction to be negligible, i.e., $\tilde{ \Delta} \simeq \Delta$. While in the numerical study we have considered two phonon modes of close frequencies, for simplicity we assume equal frequencies $\omega_{b_{(1,2)}}/2\pi \,\,{=}\,\, 10$~MHz for a brief estimation. We obtain $|\langle b \rangle| = G_0 |\langle m \rangle|^2/\omega_b \simeq 7.2 \times 10^5$, and the frequency shift $2 G_0 |\langle b \rangle| \simeq 4.5 \times 10^5$ Hz, which is much smaller than $\Delta = \omega_b \simeq 6.3 \times 10^7$ Hz, and thus can be safely neglected.

We have further adopted strong pumps for the magnon modes, which may bring in unwanted nonlinearities owing to the Kerr nonlinear term $\KK m^{\dag}m m^{\dag} m$ in the Hamiltonian~\cite{Wang}, where $\KK$ is the Kerr coefficient. For a 250-$\mu$m-diameter sphere, $\KK/2\pi \simeq 6.4$~nHz~\cite{Jie18}. In order to keep the Kerr effect negligible, $\KK |\langle m \rangle|^3 \ll \Omega$ must be guaranteed. With the parameters used in the plots in Fig.~\ref{Fig2}b and Fig.~\ref{Fig3}, we obtain a Rabi frequency $\Omega \simeq |\langle m \rangle| (\Delta^2 - g^2)/\Delta = 6.9 \times 10^{14}$~Hz (corresponding to the drive magnetic field $B_0 \simeq 3.8 \times 10^{-5}$~T and drive power $P=8.3$~mW~\cite{Note3}), and we thus have $\KK |\langle m \rangle|^3 \simeq 6.9 \times 10^{13}$~Hz $\ll \Omega$. Therefore, the Kerr nonlinearity can also be safely neglected in our linearized model.

\section{Conclusions}

We have presented a protocol to entangle the vibrational modes of two massive ferromagnetic spheres in a hybrid cavity-magnon-phonon system. The cavity-magnon subsystem has an intrinsic state-swap interaction, whereas the magnon-phonon subsystem is coupled by a nonlinear magnetostrictive interaction. We therefore directly drive the magnon mode with a red-detuned MW field to activate the magnomechanical state-swap interaction. This allows for the successive transfer of quantum correlations from a two-mode squeezed driving field to two cavity modes, then to two magnon modes, and finally to two phonon modes. We further analyze the validity of the model in detail by confirming the conditions of the approximations that have been made, and the feasibility of the protocol by considering realistic parameters, as well as experimentally accessible squeezing in MW sources. Our work studies quantum entanglement between two truly massive objects and may find applications in the study of macroscopic quantum mechanics and gravitational quantum physics.

\section*{Acknowledgments}

This work is supported by the Foundation for Fundamental Research on Matter (FOM) Projectruimte grant (16PR1054), the European Research Council (ERC StG Strong-Q, 676842), and by the Netherlands Organization for Scientific Research (NWO/OCW), as part of the Frontiers of Nanoscience program, as well as through Vidi (680-47-541/994) and Vrij Programma (680-92-18-04) grants.

\setcounter{figure}{0}
\renewcommand{\thefigure}{A\arabic{figure}}
\setcounter{equation}{0}
\renewcommand{\theequation}{A\arabic{equation}}

\section*{Appendix A:\ Drift matrix}

Here we provide the specific form of the drift matrix $A$ used in Eq.~\eqref{uAn}, which can be constructed in the form of
\begin{equation}
A=
\begin{pmatrix}
\begin{array}{c|c|c}
  A_c & A_{cm} & 0_4   \\
  \hline
  A_{cm}  & A_m & A_{mb}   \\
    \hline
  0_4 & -A_{mb}  & A_b   \\
\end{array}
\end{pmatrix} ,
\end{equation}
where $0_4$ is the $4 \times 4$ zero matrix, $A_c = - {\rm diag} ( \kappa _{a_1}, \kappa _{a_1} , \kappa _{a_2}, \kappa _{a_2})$, $A_m = - {\rm diag} ( \kappa _{m_1}, \kappa _{m_1} , \kappa _{m_2}, \kappa _{m_2})$, $A_b = - {\rm diag} ( \gamma _1, \gamma _1, \gamma _2, \gamma _2)$, and $A_{cm}$ and $A_{mb}$ are the coupling matrices for the cavity-magnon and magnon-phonon subsystems, respectively, which are given by
\begin{equation}
A_{cm} =
\begin{pmatrix}
0 &  g_1  &  0  &  0   \\
-g_1  & 0  &  0  &  0   \\
0 &  0  &  0  &  g_2    \\
0 &  0  &  -g_2  &  0   \\
\end{pmatrix} ,
\end{equation}
and $A_{mb} = - {\rm diag} ( G_1, G_1, G_2, G_2)$.

\section*{Appendix B:\ Entanglement measure -- logarithmic negativity}

The entanglement of two-mode Gaussian states can be quantified by the logarithmic negativity~\cite{LogNeg}, which is defined as~\cite{Adesso}
\begin{equation}
E_N := \max \big[ 0, \, -\ln2\tilde\nu_- \big] ,
\end{equation}
where $\tilde\nu_-\,\,{=}\,\min{\rm eig}|i\Omega_2\tilde{\CC}_{b}|$ (with the symplectic matrix $\Omega_2=\oplus^2_{j=1} \! i\sigma_y$ and the $y$-Pauli matrix $\sigma_y$) is the minimum symplectic eigenvalue of the CM $\tilde{\CC}_{b}=\PP \CC_{b} \PP$, with $\CC_{b}$ the CM of two phonon modes, which is obtained by removing in $\CC$ the rows and columns related to the cavity and magnon modes, and $\PP={\rm diag}(1,-1,1,1)$ is the matrix that performs partial transposition on CMs~\cite{Simon}. In the same way, we can calculate the logarithmic negativity of two cavity/magnon modes.


\begin{thebibliography}{99}
\bibitem{Note}
Our model is, however, not limited to this special case, because the QLEs~\eqref{QLE3}-\eqref{uAn} are derived by assuming only that, {\it in each cavity}, the cavity and magnon modes are resonant with the squeezed drive field and the blue mechanical sideband, i.e., $\omega_{aj}=\omega_{mj}=\omega_{sj}$, and $\Delta_j \simeq \omega_{bj}$ ($j=1,2$).
\end{thebibliography}
\end{document}